\documentclass[%
reprint,
superscriptaddress,
 amsmath,amssymb,
 aps,
 prl,
]{revtex4-1}
\pdfoutput=1
\usepackage{mathptmx}

\usepackage[squaren]{SIunits}  
\usepackage{graphicx}    
\usepackage{color}

\usepackage{hyperref}

\newcommand{\bra}[1]{\langle #1|}
\newcommand{\ket}[1]{|#1\nolinebreak[4]\rangle}

\begin{document}

%\preprint{}

\title{Ensemble based quantum metrology}


%\author{}
%\thanks{}
%\affiliation{}


\author{Marcus Schaffry}
\author{Erik M. Gauger}
\affiliation{Department of Materials, University of Oxford, Parks Road, Oxford OX1 3PH, United Kingdom}
\author{John J. L. Morton}
\affiliation{Department of Materials, University of Oxford, Parks Road, Oxford OX1 3PH, United Kingdom}
\affiliation{CAESR, The Clarendon Laboratory, Department of Physics, University of Oxford, OX1 3PU, United Kingdom}
\author{Joseph Fitzsimons}
\affiliation{Department of Materials, University of Oxford, Parks Road, Oxford OX1 3PH, United Kingdom}
\affiliation{Institute for Quantum Computing, University of Waterloo, Waterloo, Ontario, Canada}
\author{Simon C. Benjamin}
\affiliation{Department of Materials, University of Oxford, Parks Road, Oxford OX1 3PH, United Kingdom}
\affiliation{Centre for Quantum Technologies, National University of Singapore, 3 Science Drive 2, Singapore 117543}
\author{Brendon W. Lovett}
\affiliation{Department of Materials, University of Oxford, Parks Road, Oxford OX1 3PH, United Kingdom}

\date{\today}

%% Allowed are 600 characters including spaces
\begin{abstract}
The field of quantum metrology promises measurement devices that are fundamentally superior to conventional technologies. Specifically, when quantum entanglement is harnessed the precision achieved is supposed to scale more favourably with the resources employed, such as system size and the time required. Here we consider measurement of magnetic field strength using an ensemble of spins, and we identify a third essential resource: the initial system polarisation, i.e.\ the low entropy of the original state. We find that performance depends crucially on the form of decoherence present; for a plausible dephasing model, we describe a quantum strategy which can indeed beat the standard quantum limit.
\end{abstract}

%\pacs{}
%\keywords{}

\maketitle

Quantum metrology deals with the physical limits to measurement. Typically one prepares a probe system of size $K$ in a suitable initial state, this system acquires information about the quantity of interest, and then the probe system is measured. The process may be repeated either with a series of probes over time or, as in the present analysis, with multiple probe systems simultaneously.  The way in which the probe is prepared is closely related to the uncertainty of the parameter estimation: if prepared in a non-entangled state then the minimal uncertainty achievable scales with $1/\sqrt{K}$ \cite{Pezze.Smerzi2009EntanglementNonlinearDynamics}, the so-called Standard Quantum Limit (SQL). However, this limit can be beaten if we allow arbitrary states for the preparation of the probe system, i.e.\ if we include entangled states, as demonstrated in recent experiments \cite{Leibfried.Knill.ea2005Creationofsix-atom,Nagata.Okamoto.ea2007BeatingStandardQuantum,Pezze.Smerzi2008Mach-ZehnderInterferometryat,Afek.Ambar.ea2010High-NOONStatesby}. In idealised cases, the minimal uncertainty achievable scales with $1/K$ -- the Heisenberg limit \cite{Giovannetti.Lloyd.ea2004Quantum-EnhancedMeasurementsBeating,*Giovannetti.Lloyd.ea2006QuantumMetrology} -- which can be achieved by making use of GHZ-states, also called NOON-states \cite{Lee.Kok.ea2002quantumRosettastone}.

In optical quantum metrology the probe system is a particular state of $K$ photons, for example a NOON-state, which is a superposition of `all $K$ photons in channel $A$' with `all photons in channel $B$'. When an optical element inducing an unknown phase shift is placed in channel $A$, then the probe acquires an internal phase $K$ times as great as that which would be acquired by a single photon, and information about this phase is measured through an interference effect. In the present analysis we consider an analogous experiment involving $K$ atomic spins in a large molecule, which probes the strength of an external magnetic field. An essential difference is that we consider a large ensemble of probe molecules which are necessarily prepared, exposed to the field, and ultimately measured {\em collectively}~\cite{Jones.Karlen.ea2009MagneticFieldSensing,Simmons.Jones.ea2009Magneticfieldsensors} -- that is, addressing of individual probe molecules is impossible.

The dynamics of an ensemble is typically observed by measuring the free induction decay (FID) spin signal. Monitoring the FID can be seen as a continuous and simultaneous measurement of two non-commuting observables \cite{Esteve.Raimond.ea2004QuantumEntanglementand}. Here the observed system is barely altered by the measurement, as the number of spins in a typical sample is usually so large. Given this type of measurement, no scaling laws for the precision have been reported yet. Recent studies have rather looked at the effect of temperature on the Fisher information of three types of states \cite{Modi.Williamson.ea2010Entanglementversusclassical} and waveform estimation and its implications for quantum sensing \cite{Tsang.Caves2010FundamentalQuantumLimit}.

\begin{figure}[hbtp]
  \centering
  \includegraphics{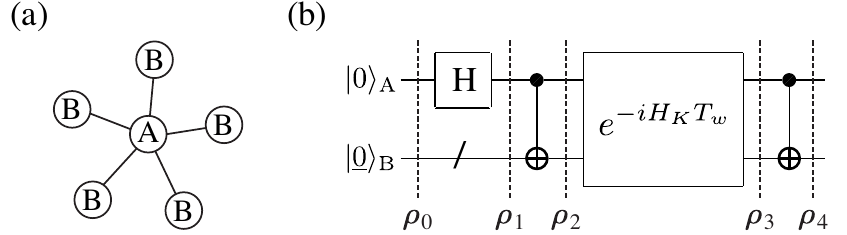}
 \caption{(a) Schematic of a sensor molecule with five satellites. (b) Quantum circuit employed in the quantum strategy.}
  \label{fig:star-topology}
\end{figure}

In this Letter we compare two strategies for measuring a small shift $\delta\!B$ of a probe magnetic field from a reference field in a spin ensemble setting \cite{Jones.Karlen.ea2009MagneticFieldSensing,Simmons.Jones.ea2009Magneticfieldsensors}. This problem is equivalent to measuring the Larmor frequency $\delta =\tfrac{\delta \! B}{\gamma}$ of a precessing spin in the probe field, where $\gamma$ is the gyromagnetic ratio. In the \emph{quantum} strategy, we consider a macroscopic ensemble of $N$ sensor molecules each consisting of $K$ spins where each molecule is prepared in a very sensitive entangled state (see Fig. \ref{fig:star-topology}). This state senses the field for a wait time $T_w$ by acquiring a  phase $K \delta T_w$. We then map the phase onto one spin of the molecule from which it can be read-out by observing the free induction decay (FID). As a performance benchmark, we compare this to the \emph{classical} or \emph{standard} strategy, where we determine the Larmor frequency by observing the FID of the same number of uncoupled spins. We will focus on the resources consumed by the two strategies and on the impact of decoherence.

The role of decoherence is very non-trivial. In the absence of decoherence and for a perfect projective measurement, the lowest achievable uncertainty in estimating $\delta$ is $1/K\sqrt{N}$ for the quantum strategy and $1/\sqrt{KN}$ for the classical strategy \cite{Giovannetti.Lloyd.ea2004Quantum-EnhancedMeasurementsBeating,*Giovannetti.Lloyd.ea2006QuantumMetrology}. Hence, the discrepancy in the precision of the two strategies increases with the size of the entangled state $K$. However, the spins will be subject to decoherence in any real-world experiment, and the sensitive entangled states will decohere faster than separable states when $K$ becomes larger \cite{Shaji.Caves2007Qubitmetrologyand}. This increased effective decoherence rate competes with the enhanced precision, making it difficult to predict the performance of the quantum strategy.

In the following, we shall determine lower bounds for the uncertainty of the parameter estimation from a measured FID for both strategies. First, we must generate a suitable entangled (GHZ) state to obtain enhanced sensitivity with the quantum strategy, meaning we require some amount of quantum control over the molecules. The details of how this is accomplished are unimportant and we shall in following consider the example shown in Fig. \ref{fig:star-topology}; molecules with this star topology have been employed in recent experiments \cite{Jones.Karlen.ea2009MagneticFieldSensing,Simmons.Jones.ea2009Magneticfieldsensors}. Each molecule consists of one central spin A and $K-1$ non-interacting satellites of type B. The satellites interact with the central spin through an Ising type interaction, leading to the following Hamiltonian in an external magnetic field (in natural units, i.e.\ $\hbar=1$):
\begin{equation}
  H_K= \gamma \delta \sigma_z^A + \delta \sum_{j=1}^{K-1} \sigma_z^{B_j} + J \sum_{j=1}^{K-1} \sigma_z^A \otimes \sigma_z^{B_j} \quad,
\end{equation}
where $\sigma_z$ denotes the usual Pauli matrix with eigenvalues $\pm \frac{1}{2}$. $\delta$ denotes the Zeeman splitting of the B spins and $\gamma = \frac{\gamma_A}{\gamma_B}$ the ratio of the gyromagnetic numbers of A and B, and we assume $\gamma \approx 1$. $J$ describes the Ising interaction strength. $\delta$ is the parameter to be estimated.

\begin{figure}[hbtp]
  \centering
  \includegraphics{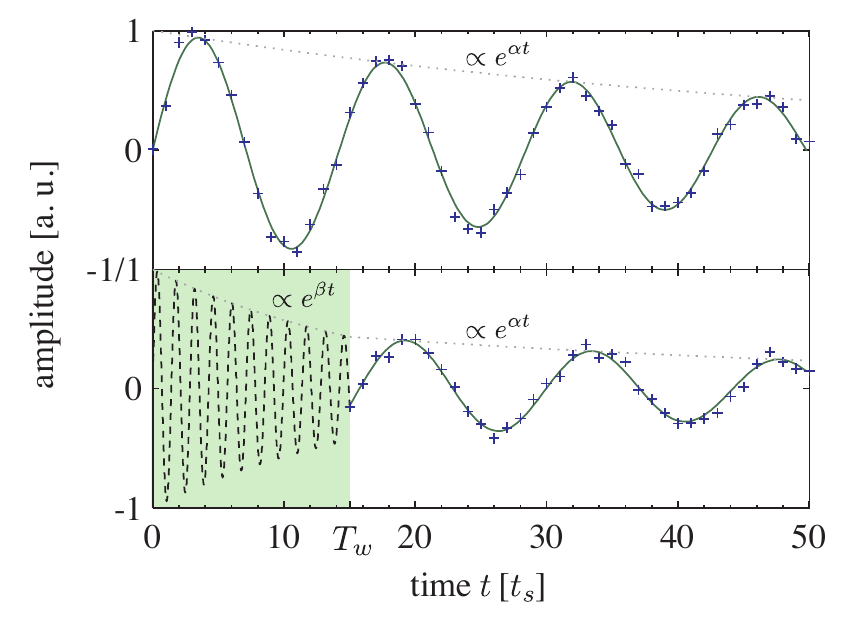}
 \caption{(color online) Simulation of the free induction decay (FID) of the classical (upper) and quantum strategy (lower); see Eqs.~(\eqref{eq:FID-CC}) and (\eqref{eq:FID-QC}). The simulated measurement points (crosses) were chosen randomly from a Gaussian distribution with a standard deviation of $0.05$. In the classical strategy we observe uncoupled precessing spins and the strength of the probe field is given by the oscillation frequency. In the quantum strategy, a GHZ-state senses the field for a time $T_w$ without producing a signal, but acquiring phase $K$ times faster than a single spin and dephasing at a rate $\beta$. The acquired phase is mapped onto the central spin and the FID is measured. The strength of the magnetic field can be now estimated from the phase and the oscillation frequency.}
  \label{fig:FID}
\end{figure}

We focus on the fundamental comparison between classical and quantum strategy given a fully polarized initial state of the sample:
\begin{equation}
\rho_0= \ket{0}_A\bra{0}_A \otimes \ket{0\ldots0}_B\bra{0\ldots0}_B = \ket{0 \underline{0}}\bra{0 \underline{0}} \quad,
\end{equation}
where the underscore denotes the state of the $K-1$ satellite spin register. For the example molecule shown, the GHZ-state can be constructed from this initial state by applying the pulse sequence shown in Fig. \ref{fig:star-topology}, i.e.\ first a Hadamard gate on the central spin and then a CNOT gate (control qubit = central spin, target qubits = satellites). The resulting GHZ-state $\tfrac{1}{\sqrt{2}}\bigl(\ket{0 \underline{0}}+\ket{1 \underline{1}}\bigr)$ freely evolves for a time $T_w$, acquiring phase $K$ times faster than a single spin. However, at the same time it is also more vulnerable to decoherence.  In practice, the dephasing rate of an individual spin, $\alpha =-\tfrac{1}{T_2^*}<0$, is limited by inhomogeneous broadening~\cite{Levitt2002SpindynamicsBasics}, thus we can neglect spin flip processes. Of course the dephasing rate of the GHZ-state $\beta<0$ is related to $\alpha$, and we shall discuss this dependence later. The state of the system after the wait time $T_{w}$ is then:
\begin{align}
\rho_3 = \frac{1}{2} \Bigl( \ket{0\underline{0}}\bra{0\underline{0}} +  \ket{1\underline{1}}\bra{1\underline{1}} &+ e^{-K \delta T_w i+\beta T_w} \ket{0\underline{0}}\bra{1\underline{1}} +\nonumber\\
&+e^{K\delta T_w i +\beta T_w}  \ket{1\underline{1}}\bra{0\underline{0}} \Bigr) \quad.
\end{align}
To measure the acquired phase $K \delta T_w$ we map the GHZ-state onto the central spins by applying a CNOT gate. These spins are then measured by observing the decay of the transverse magnetization at $M \in \mathbb{N}$ discrete points in time, separated by the sampling time $t_s$. The observed signal $x_m$ at time $mt_s$ ($m=0,\ldots,M-1$) can be modelled as a sum of Gaussian distributed noise $b_m$ and the ideal signal $\hat{x}_m=\langle X+i Y \rangle(m t_s)$ \cite{Cavassila.Deval.ea2000Cramer-RaoBoundExpressions}
\begin{equation}
\label{eq:FID-QC}
 x_m   =\hat{x}_m+ b_m = c  e^{K\delta T_w i + \beta T_w } e^{\delta  m t_s + \alpha m t_s }  + b_m \quad ,
\end{equation}
where $c$ is a proportionality factor that depends on the number of molecules in the sample. A simulation of the FID is shown in Fig. \ref{fig:FID}.

A suitable metric for the precision of the measurement is the Cram\'{e}r-Rao bound (CRB) \cite{Cavassila.Deval.ea2000Cramer-RaoBoundExpressions}, which essentially offers a lower bound on the uncertainty (standard deviation) $\sigma_{p_{\ell}} $ of an estimated parameter $p_\ell$ ($\ell=1,\ldots,P$):
\begin{equation}
 \sigma_{p_{\ell}} \geq \text{CRB}_{p_{\ell}} = \sqrt{(F^{-1})_{\ell \ell}} \quad.
\end{equation}
Here, $F$ denotes the Fisher information \cite{Bos1982HandbookofMeasurement} given by the real part of a complex-valued matrix product
\begin{equation}
 F= \frac{1}{\sigma^2} \Re( D^{\dagger} D ) \quad,
\end{equation}
where $D_{ij}=\frac{\partial \hat{x}_i}{\partial p_j}$ and the partial derivatives are evaluated at the parameters that we are going to estimate. In our case the parameters are $c, \alpha,$ and $\delta$. $\sigma=\sigma_r=\sigma_i$ denotes the standard deviation of the real/imaginary part of the noise. Inverting the 3x3 Fisher information matrix we obtain:
\begin{equation}
\label{eq:CRB-GHZ}
\text{CRB}_{\delta,\text{GHZ}} =   \frac{ e^{-\beta T_w} }{\sqrt{\sum\limits_{m=0}^{M-1} \bigl( K T_w  + m t_s \bigr)^2 \exp(2 \alpha m t_s)}} \frac{\sigma}{c} \quad.
\end{equation}

Next we determine the CRB for the classical strategy. Here we are given $N \cdot K$ identical and uncoupled spins. We obtain the relevant signal by rotating the initial state $\rho_0$ into the $xy$-plane with a global Hadamard gate followed by measuring the transverse magnetization. The resulting density matrix evolves as 
\begin{equation}
    \rho_{4}(t) = \frac{1}{2}\begin{pmatrix} 1 &  e^{i \delta t+\alpha t}  \\ e^{- i \delta t+\alpha t} & 1 \end{pmatrix} \quad,
\end{equation}
giving rise to a measured signal of the following form
\begin{equation}
\label{eq:FID-CC}
  x_m'= c'  e^{ \delta   m t_s i+\alpha m t_s } + b_m' \quad.
\end{equation}
Analogously to above we obtain the CRB for this signal:
\begin{equation}
\label{eq:CRB-STD}
 \text{CRB}_{\delta,\text{STD}} = \frac{1}{\sqrt{\sum\limits_{m=0}^{M'-1} (m t_s)^2 \exp(2 \alpha m t_s)}} \frac{\sigma'}{c'} \quad.
\end{equation}
This expression is a lower bound on the uncertainty for independent spins against which we shall benchmark the quantum strategy. Hence it may be considered as the SQL in this setting of spin ensemble measurements. For a high signal-to-noise ratio (SNR) achieved by a sufficiently large number of molecules $N$, there exists an efficient estimator  which matches the accuracy predicted by the CRB  \cite{Bresler.Macovski1986Exactmaximumlikelihood}. Hence, we can directly use Equations \eqref{eq:CRB-GHZ} and \eqref{eq:CRB-STD} to compare the two strategies.

First, we need to specify a fair comparison with the same resource allocation for both strategies. For conventional quantum metrology with projective measurements the challenging question of a fair resource comparison has recently been addressed in Ref.~\cite{Zwierz.Perez-Delgado.ea2010Generaloptimalityof}. In the present case of ensemble quantum metrology, we shall at first allow both strategies to observe the full FID while consuming the von Neumann entropy of $1\cdot N$ spins. As we have defined it, the quantum strategy consumes one spin per molecule (the central spin A) in the measurement and all other spins remain pure. This implies that we must also only measure $1\cdot N$ spins instead of $K\cdot N$ spins for the classical strategy, meaning that the SNR of both measurements are equal $\frac{c'}{\sigma'} = \frac{c}{\sigma}$.

At first sight this way of counting resources may look biased towards the quantum case, as we do not seem to take the satellite spins into account. Nonetheless, our comparison is fair: the $K-1$ satellite spins act as an antenna to pick up phase more rapidly, yet they are not `consumed' (this is a direct consequence of the dephasing model of decoherence). After the quantum strategy is complete, the central spin is measured (its polarisation is lost) but all satellite spins are back in the pure state $\ket{\underline{0}}$ and could be recycled to obtain a further parameter estimate. The accuracy of such an estimate, made using any sensible protocol on these remaining $(K-1)N$ spins, can be no worse than that obtained using the standard strategy. This observation validates the classical resource count stated in the previous paragraph.

The ratio of Eq. (\ref{eq:CRB-STD}) and Eq. (\ref{eq:CRB-GHZ}) can be approximated by an integral expression (assuming the sampling rate resolves the decay, i.e.~$t_s \ll T_2^*$):
\begin{align}
\label{eq:maxR-inf}
R_{\infty}:&= \frac{\text{CRB}_{\delta,\text{STD}} }{\text{CRB}_{\delta,\text{GHZ}} }\approx  \sqrt{\frac{\int_{0}^{\infty} ( K T_w  + t)^2 \exp(2 \alpha t) \,  dt}{e^{-2\beta T_w}\int_{0}^{\infty} t^2 \exp(2 \alpha t) \, dt}} \\
&= e^{\beta T_w}\sqrt{1-2  K \alpha T_w(1-K \alpha T_w)} \quad.
\end{align}
Whenever $R_{\infty}>1$ the quantum strategy outperforms the classical strategy with respect to the precision of the parameter estimation.

So far we have not yet specified how the dephasing rate $\beta$ of the GHZ-state relates to the dephasing rate $\alpha$ of a single spin, and we shall now discuss a number of different decoherence models for the GHZ-state. 

First, we consider $\beta=\alpha$, implying that the GHZ-state does not decohere faster than a single spin; this is expected for a given macroscopic magnetic field inhomogeneity. In this case, there is for any $K>1$ a wait time $T_w$ for which the quantum strategy surpasses the classical one. Conversely, completely correlated or collective noise over each molecule has the most aggressive effect on the GHZ-state \cite{Shimizu.Miyadera2002StabilityofQuantum,*Palma.Suominen.ea1996QuantumComputersand}. Here the noise can be described with a single Lindblad operator $\sum_{j=1}^K\sigma_z$ and $\beta = K^2 \alpha$, while  for uncorrelated noise $\beta= K \alpha$ \cite{Aolita.Chaves.ea2008ScalingLawsDecay}. In general we consider a power law dependence of $\beta$ on $\alpha$, i.e.\ $\beta= K^p \alpha$, where $0\leq p\leq 2$. In recent experiments the decoherence rates for highly correlated solid-state spin states were obtained experimentally \cite{Krojanski.Suter2004ScalingofDecoherence,*Krojanski.Suter2006ReducedDecoherencein}, revealing $p \approx 1/2$. The authors attributed this to non-Markovian correlated noise. A significantly smaller value for $p$ was found in \cite{Simmons.Jones.ea2009Magneticfieldsensors}, where the $T_2^*$ time of a single spin was determined to be \unit{0.37}{\second} and that of a 13 particle GHZ-state to be \unit{0.28}{\second}, which can be interpreted as a factor of $p\approx 0.11$.

\begin{figure}[hbt]
  \centering
  \includegraphics{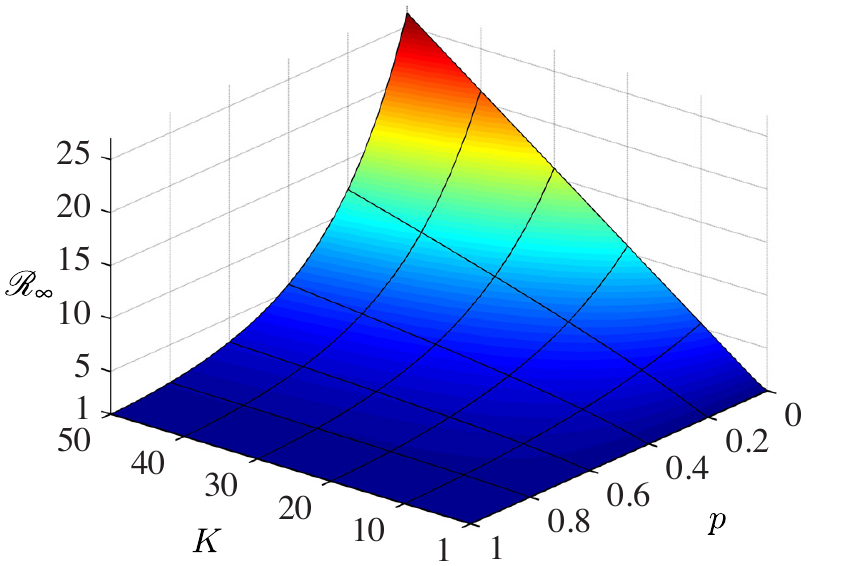}
 \caption{(color online) The maximal $\mathcal{R}_{\infty}$ for a given $K$ as a function of $K$ and $p$. $\mathcal{R}_{\infty}>1$ implies that the quantum strategy outperforms the standard strategy.}
  \label{fig:optimal}
\end{figure}

The performance of the quantum sensor depends critically on the value of $p$. One can easily check that for $1\leq p \leq 2$, $R_{\infty}< 1$ for any $T_w>0$ and $R_{\infty}=1$ only for $T_w=0$. We therefore conclude that the precision of the quantum method never outperforms the precision of the standard method, if $1\leq p \leq 2$. In contrast, for $p<1$ there is an optimal wait time for $T_w$ the GHZ state which is non-zero; this varies with $K$ but always gives $R_{\infty}>1$. Specifically, calculating the optimal waiting time $T_{w}$ that gives rise to a maximum of $\mathcal{R}_{\infty}$ (for a given $K$ and $p<1$) yields 
\begin{equation}
   \mathcal{R}_{\infty}= \max_{T_w} R_{\infty} =\frac{K^{\frac{1}{2}-p} \sqrt{ K+\sqrt{K^2-K^{2 p}}}}{\exp\left(\frac{1}{2} \left(1-K^{p-1}+\sqrt{1-K^{2 p-2}} \right)\right)} \quad,
\end{equation}
which scales like $\frac{\sqrt{2}}{e}K^{p-1}$  to leading order (see Fig. \ref{fig:optimal}). 
%The optimal waiting time is given by:
%\begin{equation}
%T_w^{\text{opt}} = \frac{1}{2} \left(K^{-p}-K^{-1}+\sqrt{K^{-2 p}-K^{-2}}\right) T_2^* \quad.
%\end{equation}
Therefore the ensemble SQL can indeed be beaten with a quantum strategy if the decoherence of the GHZ-state is not too aggressive, i.e.~for $p<1$. Moreover, we see that under this condition, the precision of the estimation improves monotonically as $K$ increases.

\begin{figure}[hbt]
  \centering
  \includegraphics{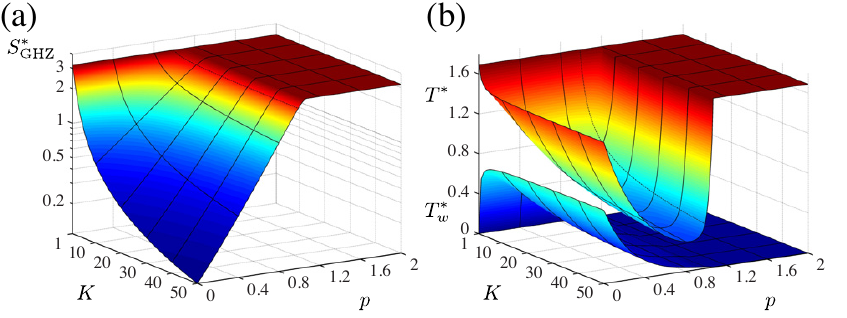}
\caption{(color online) (a) Minimal (optimized over $T$ and $T_w$) uncertainty of the parameter estimation for the quantum strategy per $\sqrt{\text{Hz}}$, i.e.\ $S_{\text{GHZ}}^*$ in units of $\sqrt{t_s} |\alpha|^{3/2}\frac{\sigma}{c}$ in dependence of $K$ and $p$. (b) The corresponding optimal times $T^*$ (upper surface) and $T_w^*$ (lower surface) in units of $T_2^*$ for which the minimum in (a) is attained in dependence of $K$ and $p$.} \label{fig:OptimalCRBGHZrtHz}
\end{figure}

So far we have neglected time as a resource, focusing instead on system size and the consumption of initial polarisation (maximising the von Neumann entropy of one spin per molecule). It is interesting to extend our analysis to a restricted process time, meaning only a part of the FID can be observed. Since the first part of the FID contains most information, this would enable a `better' sensor by a series of repeated runs in a given time window $T_{\text{tot}}$ if one had the ability to reset the spins to their initial state after time $T$, for example with an optical switch. We assume that this can be done instantly and then the uncertainty of the parameter estimation of this series is given by
\begin{equation}
   \frac{1}{\sqrt{T_{\text{tot}}/T}}\text{CRB}_{\text{STD}/\text{GHZ}}  =: \frac{1}{\sqrt{T_{\text{tot}}}} S_{\text{STD/GHZ}} \quad.
\end{equation}
Obviously one would choose the length of a time slice $T$ optimally, i.e.\ in such a way that the uncertainty per $\sqrt{\text{Hz}}$, i.e.\ $S_{\text{STD/GHZ}}$ is minimal. For the classical strategy we find by using the integral approximation for the CRB from above and numerical optimization that
\begin{equation}
  S^*_{\text{STD}} :=\min_T S_{\text{STD}} \approx 3.21 \sqrt{t_s} |\alpha|^{3/2}\frac{\sigma'}{c'} \quad,
\end{equation}
which is attained for the optimal time $T^* \approx 1.69 T_2^*$. In the quantum strategy $S_{\text{GHZ}}$ also depends on $T_w$, the amount of time for which the spins are in the GHZ-state. In contrast to the classical strategy, the minimum here depends on the system parameters $K$ and $p$. We have not been able to find an analytic expression for $S^*_{\text{GHZ}}  :=\min_{T,T_w} S_{\text{GHZ}}$ and therefore we performed a numerical optimization with the results displayed in Fig. \ref{fig:OptimalCRBGHZrtHz}. As in our first comparison we assume that the SNR for both strategies are equal. If aggressive noise is affecting the GHZ-state the optimal quantum strategy is basically the optimal standard strategy, as $T^*\rightarrow 1.69 T_2^*$ and $T^*_w\rightarrow 0$, when $p\rightarrow 2$. For small $p$ and large $K$ however the quantum strategy significantly outperforms the standard strategy. Interestingly the quantum strategy can now beat the optimized standard strategy for values $p$ that are slightly larger than one.

In conclusion we have presented a framework for analysing the performance of quantum metrology using spin ensembles. This framework incorporates the special nature of the non-projective measurement process, and leads one to consider the polarisation (i.e.\ the low entropy) of the initial state as a resource. We find that the decoherence model plays a defining role in this framework, and we have identified the parameter regime where a certain quantum strategy can beat the best classical strategy.

\begin{acknowledgments}
\textit{Acknowledgements -} We thank Y. Matsuzaki and S. Simmons for discussions. This work was supported by the EPSRC through QIP IRC (GR/S82176/01 and GR/S15808/01), the National Research Foundation and Ministry of Education, Singapore, the DAAD, and the Royal Society.
\end{acknowledgments}

%\bibliography{EnsembleBasedQuantumMetrology}

%merlin.mbs 2010-03-15 4.21a (PWD, AO, DPC)

\end{document}